\documentstyle[psfig]{mn}

\input{epsf}

\def\lsim{\mathrel{\hbox{\rlap{\hbox{\lower4pt\hbox{$\sim$}}}\hbox{$<$}}}}
\def\gsim{\mathrel{\hbox{\rlap{\hbox{\lower4pt\hbox{$\sim$}}}\hbox{$>$}}}}

\def\and   {\rm {et al.} \rm}  
\def\etal  {\rm {et al.} \rm}

\begin{document}

\title[Anisotropies in the redshift-space correlations II: Observations]
{Anisotropies in the redshift-space correlations 
of galaxy groups and clusters II: Analysis of observational cluster samples.}

\author[N. D. Padilla \& D. G. Lambas]
{N. D. Padilla$^{1}$ \& D. G. Lambas$^{2,3}$ \\
1.University of Durham, South Road, Durham DH1 3LE, UK\\
2.IATE, Observatorio Astron\'omico de C\'ordoba, Laprida 854, 5000, C\'ordoba,
Argentina\\
3.John Simon Guggenheim Fellow\\
\\
{\rm High resolution plots and info at http://star-www.dur.ac.uk/$\sim$nelsonp/anisotropies}\\
}

\maketitle 
 
\begin{abstract}
We study the redshift space  correlation function of galaxy clusters 
for observational samples constructed in different surveys. We explore correlation amplitudes,
pairwise velocity distributions and bias factors. 
Systematics in cluster identification procedures are the main source
of biased estimates of the correlation amplitude and inferred velocity dispersions.
We find that the large elongations  along the line of sight in the Abell catalogue
cannot be explained solely in terms of the  
errors in distance measurement originating
from using a small number of galaxies.  The inclusion
of a significant fraction of galaxies and systems not physically bound
 to the clusters are responsible for this large systematic effect.
We also find a significant dependence of the redshift space distortion of the
correlation function on the cluster Bautz-Morgan type, an effect that may rely
on the fact that, due to the regular appearance of low Bautz-Morgan type
clusters, a sample of such objects would 
be less contaminated. 
We confirm that the effect of a low  number of
redshift measurements, $n_z$, is to increase 
the redshift-space correlation length
and bias factor. The results are very stable for $n_z > 10$.
We also test the effects of different $n_z$ in 
the catalogue of groups/clusters 
derived from the Updated Zwicky Catalogue and find that, 
for cluster samples identified in redshift surveys, even 
a low number of redshifts $n_z \sim 5$, 
is sufficient to provide reliable results.  
By comparing our results with those of numerical simulations 
we explore the strong 
influence on the clustering distortion pattern in redshift space
from effects associated with the cluster identification
procedure from two dimensional surveys. 
The identification of clusters in X-ray surveys improves this situation, although
there are still systematic effects which are probably due to identification
of optical sources in the determination of cluster redshifts. 
These systematics are particularly strong for the most luminous X-ray selected
clusters in the Extended Bright Cluster Survey, which exhibits very large 
anisotropies, comparable to those present in the Abell catalogue.
Our results demonstrate that 
forthcoming large redshift surveys will be extremely important 
for the construction of new samples of groups and
clusters as well as improving the determination of optical and X-ray selected
cluster distances, essential for reliable analyses 
of the large scale structure. 
\end{abstract}


\section{Introduction}

Clusters of galaxies can be used as  
tracers of the large scale structure of the Universe.
However,		
their identification is not free from systematics, for example
the Abell catalogue is known to be afflicted by projection effects 
(Sutherland 1988; Sutherland \& Efstathiou 1991; see also Lucey 1983),
which are somewhat reduced when restricting the sample of clusters
to high richness class (Miller et al. 1999).
 
Cluster peculiar motions
produce an apparent
distortion of the clustering pattern as measured by the
two-point correlation function in redshift space, $\xi(\sigma,\pi)$
(Croft \& Efstathiou 1994; 
Bahcall, Cen \& Gramann 1994),
with $\sigma$ and $\pi$ being the separations perpendicular and
parallel to the line of sight, respectively.
The non-linear virialized regions induce 
elongations along the line of sight which make it possible to
estimate the
one-dimensional pairwise rms velocity dispersion, $<w_{12}^2>^{1/2}$
(Davis \& Peebles 1983).  
The infall onto overdense regions in the form of bulk motions 
dominates at large scales resulting in a compression of the $\xi$ contours
along the direction of the line of sight.

The presence of	 
spurious large anisotropies along the line of sight of
Abell clusters has been largely discussed 
(e.g., Sutherland 1988, Sutherland \& Efstathiou 1991, 
Bahcall \& Soneira 1983; Postman, Huchra \& Geller 1992).
Arising mainly from inhomogeneities  
in the detection of clusters in two dimensions (ie. prone to
strong projection effects), these observed
large anisotropies are extremely unlikely
to exist for systems of galaxies within a hierarchical scenario of 
structure formation.
Therefore, the clustering signal along the line of sight 
in the Abell catalogue could be artificially enhanced 
(Sutherland 1988; Sutherland \& Efstathiou 1991; see also Lucey 1983)
due to significant projection effects 
which could bias the determination of the 
correlation function as well as
the mass function of these systems.  
A velocity broadening of 
$2000{\rm kms}^{-1}$, 
obtained by Bahcall, Soneira \& Burgett (1986) from their analysis of the
clustering of Abell clusters, was
interpreted as arising from a combination 
of cluster
peculiar motions as well as geometrical distortions of superclusters.  
However, van Haarlem \etal (1997) find that a third of Abell clusters of
richness class $R \geq 1$ would not be real
physically bound systems but simply projections of galaxies and groups along
the line of sight.  
It has to be recalled that early surveys of Abell clusters  
contain large fractions of low richness clusters 
(Abell richness class $R=0$), which were not intended to form complete samples 
suitable for statistical analyses.  Miller \etal (1999) analysed samples of 
Abell clusters with richness $R \ge 1$, with new accurate determinations of 
cluster positions using several galaxy redshifts.  These authors find that 
the clustering signal along the line of sight is greatly reduced  
when compared with the Bahcall \& Soneira (1983) results.
The anisotropy is further reduced after the orientation of two 
superclusters that are elongated along the line of 
sight is changed. 
More recently, Miller et al. (2002) found that the inclusion
of clusters in superclusters has a significant incidence in
the anisotropies of the redshift-space correlation function.
A possible interpretation of this result is that cluster identification
in overdense regions is likely to be more affected by  projection
effects.
Peacock \& West (1992) also found that restricting attention to higher 
richness Abell clusters removed the strong radial anisotropy seen 
in the clustering measured from earlier surveys.

A more recent generation
of cluster catalogues presumably  less affected
by identification biases (APM: Dalton \etal 1992, 1994, 1997; 
Cosmos: Lumsden \etal 1992) was  
drawn from machine-scanned survey plates with better 
calibrated photometry.
With the aim of reducing biases in cluster richness determinations
arising from projection effects, the 
search radius used to define clusters in these machine 
based catalogues is smaller than that used by Abell.
As expected, the clustering pattern
found in these more recent cluster redshift surveys 
does not exhibit very large enhancements along the line of sight; furthermore, 
the trend of increasing correlation amplitude with decreasing space 
density of clusters is weaker than that found for Abell clusters 
(Croft \etal 1997) and is similar to the trend expected in current models,
where there is only a weak dependence of correlation 
length on cluster space density, as is obtained in other all sky X-ray surveys 
(eg REFLEX sample, Collins et al. 2000), 
and samples of Abell clusters confirmed in X-rays
(X-ray Bright Abell Cluster Sample, XBACS, Ebeling et al. 1996). 
For these samples, systematic projection effects along the line of sight 
are expected to be significantly lower than in the Abell catalogue 
(Abadi, Lambas \& Muriel 1998; Borgani et al. 1999; Collins et al. 2000).
However,  it is important to consider that
the analysis of high richness Abell clusters, performed by Miller
\etal (1999), gives results comparable to the amount of distortion of the 
clustering pattern found for APM clusters (see Fig 5 of Miller \etal 1999).

In this paper, we use the results
of an analysis of the redshift space clustering of massive 
dark matter haloes in mock cluster samples
(Padilla \& Lambas, 2003, Paper I hereafter) extracted from
the $\tau$CDM Hubble Volume simulation.
The results from Paper I are used in the assessment 
of projection biases on catalogues
of clusters identified in two dimensions and three dimensions.
In paper I we studied statistically the redshift-space
correlation function as a function of coordinates parallel and perpendicular 
to the line of sight, through the pairwise velocity dispersions, redshift-space 
correlation length and bias parameters obtained from different mock cluster 
samples constructed using several cluster identification algorithms, and 
different number of redshifts and cluster search radii.  
Spurious distortions in the redshift space cluster correlation function
could originate in various systematics associated to these issues. For instance
in several catalogues, many cluster distances are
determined by a single galaxy redshift, usually the brightest cluster member
and therefore are subject to projection contamination 
from background and foreground galaxies. 

In this paper we present an analysis of
the correlation function of observational
cluster data, including optical and X-ray cluster samples.  
In section 2 we present the statistical procedures adopted 
and we
summarise the results obtained from the study of
different mock samples in Paper I that concern the present study. 
These mock samples
are used for assessing the degree to which
different problems could contribute to the anisotropies detected in
observational correlation functions in Section 3. In this section 
we study the effects of the number of redshifts used in 
the determination of cluster distances, the effects arising 
from selecting samples with different Bautz-Morgan cluster types 
(Bautz \& Morgan, 1970) and the different behaviour of
samples with high and low x-ray luminosities.  
In section 4, we analyse groups of galaxies obtained from the Updated Zwicky 
Catalogue, hereafter UZC , which, in contrast to other 
sets of clusters  studied in this paper, were identified from redshift space
data.  In section 5 
we discuss our results and present the main conclusions.

\section{Statistical procedures and results from mock catalogues}

We followed the procedures described in paper I to measure
the redshift-space correlation function as a function of
the pair separation parallel and perpendicular to the line of sight,  
$\sigma$ and $\pi $ respectively.
The iso-correlation contours 
(curves of equal amplitude of $\xi(\sigma,\pi)=\xi^{fix}$)
provide a suitable
measurement of the pairwise velocities of galaxies
by comparing the measured and predicted contour
levels.
These contour levels are approximated
by the functions $r^m(\theta)$ and $r^p(\theta)$ for the
measured and predicted correlations respectively,
where $\theta$ is the polar angle measured from the direction perpendicular 
to the line of sight (See Paper I for details).

In order to make a proper comparison between  
model predictions and observations, we adopt
the real-space correlation function of the mass distribution
in a $\Lambda$CDM cosmology.  We use a mass density 
parameter $\Omega_{matter}=0.3$, vacuum energy density $\Omega_{\Lambda}=0.7$,
and a CDM shape parameter $\Gamma=0.2$ consistent
with recent CMB results (Netterfield et al. 2002).
The normalisation of the power spectrum, $\sigma_8=0.9$, is taken
from recent constraints obtained from the 2dF Galaxy Redshift
Survey (Norberg et al. 2002). Given the relatively large
 uncertainties in the cluster correlation functions the
adopted transfer function does not include baryons, however 
our choice of
$\Gamma$ implicitly allows for a combination of baryon fraction and
an effective value of $\Gamma$ as discussed in Eisenstein \& Hu (1996).  
We obtain the real-space 
correlation function by Fourier transforming the CDM power spectrum
\begin{equation}
\xi^{CDM}(r)= \frac{1}{2\pi^2} b^2 \int_0^{\infty} P(k) 
\frac{\sin(kr)}{kr} k^2 {\rm d}k,
\label{eq:xicdm}
\end{equation}
and then use this to evaluate the theoretical prediction for $\xi^p(\sigma,\pi)$. 
We search for the optimum values of the scale independent bias parameter 
$b$ and $w_{12}$ by minimising the quantity $\chi^2$,
\begin{equation} 
\chi^2 = \sum_{i} [r^m_l(\theta_i)-r^p_l(\theta_i)]^2,
\label{eq:chi}
\end{equation} 
where we have chosen to compare a set of discrete levels, $l=0.6,0.8,1.0,1.2$
and $1.4$, of the redshift-space 
correlation function $\xi(\sigma,\pi)$ amplitude instead
of comparing values of the correlation function on a grid of $\sigma$ and 
$\pi$ distances.  Our
decision is based on the fact that more reliable and stable results are 
obtained when using the isopleth comparison technique (Padilla et al. 2001).
We notice that the effect of adopting a different normalisation for
the theoretical power spectrum mainly affects the bias parameter $b$,
leaving the pairwise velocity $w_{12}$ mostly unchanged.

We briefly summarise the results obtained for
the statistical properties of cluster samples drawn from mock catalogues
presented in Paper I.  These mock
catalogues have been obtained from one of the mock APM galaxy
catalogues extracted from the $\tau$CDM Hubble Volume
simulation by the Durham Extragalactic Group (Evrard et al., 2002)
following a procedure similar to that used in Cole et al. (1998). 

In Paper I, we used three main algorithms to search for groups 
in the numerical simulations, and constructed four samples of 
clusters from the simulations.  The aim was
to understand the effects of cluster 
selection procedures on the  cluster correlation functions,
and in particular, to establish the origin of the
large elongations along the line of sight present in the
measured correlation functions.   
Sample 1, constructed using a 
friends-of-friends algorithm  applied to
the particles in the simulation box,
has a lack of a significant number of 
spurious clusters.  The measurements
of anisotropies in the correlation function clearly show
the expected infall pattern.  
Sample 2 allows to understand the results obtained from the 
identification of groups in a mock catalogue with redshift information 
following the lines described by Huchra \& Geller (1982).
In this case the results from Paper I indicate that the infall
pattern is not as readily visible as in the results from Sample 1,
but a small $n_z \simeq 5$ is enough to produce it.
Sample 3 was constructed 
following the prescription described by Lumsden et al. (1992),
using an angular search radius $r_{c}=1.0$h$^{-1}$Mpc. 
This procedure was applied to the same mock catalogue as the second algorithm,
and the resulting
sample was found to show severe projection effects. 
The results from this sample will be used 
to assess the degree of these effects
in observational samples of cluster of galaxies identified from angular
data.  
In Paper I we also extracted another sample combining the mock catalogues
defined above, in order to mimic as closely as possible
the methods used in obtaining the observational samples.  
From the clusters identified using angular data (sample 3), we drawn
a restricted set of clusters of galaxies, whose angular positions lie
within the cluster search radius from a 3D identified cluster, 
sample 4. 

As a summary of the results of our mock samples, we recall figures
11,12, and 13, of Paper I,
where the behaviour of the correlation
contours is assessed by measuring the pairwise velocity dispersion $w_{12}$ and
the correlation length $s_0$ as a function
of $n_z$.  We found that in the case of the identification
of clusters from angular data, the values of $s_0$ and $w_{12}$ are
higher as the underlying value by as much as an order of magnitude.
In the case of clusters identified from 3-dimensional information (sample 2), 
these quantities converge rapidly to their actual values, which are
obtained when a small number of galaxies, not much larger than $5$, is used
to determine the cluster distance.

For more detailed descriptions of the characteristics
of each sample, analysing the differences in elongations arising
from using different $n_z$, and search radius $r_c$,
we refer the reader to Paper I.

\section {Redshift Space Distortions of Groups and  Clusters of Galaxies}

In this section we 
compute the two-point 
correlation function in redshift space using different cluster catalogues.

Following usual procedures we estimate
$\xi^o(\sigma, \pi)$ by counting pairs in the data and in
a random catalogue with the 
same angular limits and radial selection function than the observational sample.
Data-data and random-random pairs ($N_{dd}$ and $N_{rr}$ 
respectively) are  binned as a function
of separation in the two variables $\sigma$ and $\pi$.  
We adopt the Davis \& Peebles (1983) estimator,
\begin{equation}
\xi^o(\sigma,\pi)=\frac{N_{dd} n_R^2}{N_{rr} n_D^2}-1
\end{equation}
where $n_D$ and $n_R$ are the number of data and random points respectively. 

Our analyses 
are centred on large correlation amplitude signals 
which are  not strongly sensitive to the choice of estimator
(see Hamilton 1993).

In Figure 
\ref{fig:dist.str_apm},
we show the contours levels of
$\xi^o(\sigma, \pi)$
for  UZC groups (left panel).
There are no signals
of distortions along the line of sight direction for the groups
consistent with low
pairwise velocities of groups. By comparison with the results 
for samples of clusters identified in two-dimensional surveys  
shown in the other panels of this figure
it is evident the existence of either large pairwise 
velocities or large projection biases in these samples, 
as also discussed by Sutherland (1988).
Such elongations could originate in 
the systematic presence of groups along the line of sight, in the fields
of clusters identified from angular data.
From the theoretical point of view, strong elongations along the line of
sight are not expected in a hierarchical scenario of structure formation.

This section is divided as follows:
In section \ref{ssec:optical} we study the correlation function
of APM and Abell clusters.  We divide the Abell samples according
to the number of redshifts used in the determination of the cluster
distances in section \ref{ssec:opticalnz}, and also according
to Bautz-Morgan type in section \ref{ssec:opticalbm}.
Cluster samples identified in X-rays are studied in section
\ref{ssec:xray}; the results from subdividing the EBCS
sample using a x-ray luminosity cut are shown in section
\ref{ssec:xrayluminosity}; and the outcomes of sub-dividing the
XBACS sample
according to different Bautz-Morgan type are explained in section
\ref{ssec:xraybm}.  Finally, the inferred values of relative velocities
and redshift-space correlation lengths are presented in section
\ref{ssec:w12obs}, and the bias factors in section
\ref{ssec:biasobs}.

\subsection{Clusters samples selected from angular surveys}
\label{ssec:optical}

In this section we study the iso-correlation contours obtained from
cluster catalogues identified using surveys of galaxies containing only angular positions.  As
has already been pointed out, studies of such samples have shown
large elongations along the line of sight
even when, in light of the previous discussion,
we would have expected infall patterns from samples of massive clusters
(Padilla \& Baugh, 2002).

\begin{figure*}
{\epsfxsize=17.5truecm \epsfysize=5.3truecm 
\epsfbox[31 22 856 271]{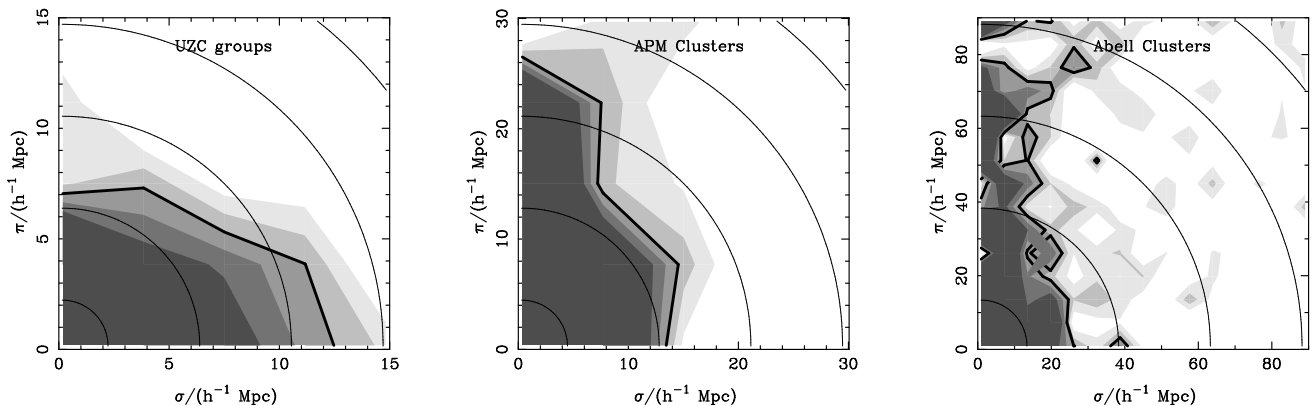}}
\caption{
Correlation function of UZC groups (left hand side panel),
APM clusters (Dalton et al. 1992; middle panel), and
the full sample of Abell clusters with richness class $r \ge 1$ and
measured redshifts (Struble \& Rood, 1999, right panel).
The transitions between different 
shadings correspond to fixed values of $\xi=0.6,0.8,1.0,1.2,$ and $1.4$ 
levels, which are the values used later to infer the relative
velocities, $w_{12}$. 
The thick line corresponds to the $\xi=1$ level, and the thin lines
show the expected contours for a spherically symmetric correlation function. 
}
\label{fig:dist.str_apm}
\end{figure*}

We measured the correlation function of APM and Abell clusters
in order to compare them to the results from the UZC groups,
shown in the left panel of
figure \ref{fig:dist.str_apm}.  We also show in this figure the results 
for APM clusters (Dalton et al., 1992; middle panel), and
Abell clusters with measured redshifts (Struble \& Rood 1999, right panel).
The sample of Abell clusters consists
only of clusters with determined Bautz-Morgan type.
It can be seen that there is a marked difference in the distortion pattern 
between the UZC groups, which show an infall signature,
and the elongated correlation functions corresponding to
the other two cluster samples in this figure.
This elongation is most severe in the
case of Abell clusters, as it has already been pointed out in several
previous works (Bahcall \& Soneira 1983; Postman, Huchra \& Geller 1992),
and has been reconciled by considering
the effects of cluster alignments in supercluster structures
(Bahcall, Soneira \& Burgett 1986), and
the likely inclusion of spurious clusters in the sample due
to projection effects 
(Sutherland 1988; Sutherland \& Efstathiou 1991; see also Lucey 1983).
APM clusters show a much milder elongation, which can be
easily reproduced in numerical simulations when a redshift error bigger
than $700$km/s is added to the mock cluster positions (Padilla \&
Baugh 2002) although contamination can not be totally excluded.

In an attempt to explore and quantify
the origin of the problem of large elongations of the clustering signal
we study the effects of errors in cluster distances 
by considering different number of redshifts used in the determination 
of cluster distances, as well as the
possible projection effects.  In order to do this, we use results
obtained from mock catalogues 
where clusters of galaxies are identified from angular data
(see Paper I).  Such a comparison will also
be useful in providing a quantitative determination of the degree of projection 
effects present in the Abell cluster sample.

\subsubsection{Effects of different number of redshifts}
\label{ssec:opticalnz}

In order to assess to which extent the use of samples of clusters
with large number of redshift
measurements, $n_z$, reduces the distortion of the clustering pattern, 
we have explored samples with different values of $n_z$.

\begin{figure*}
{\epsfxsize=15.truecm \epsfysize=8.truecm 
\epsfbox[33 322 557 561]{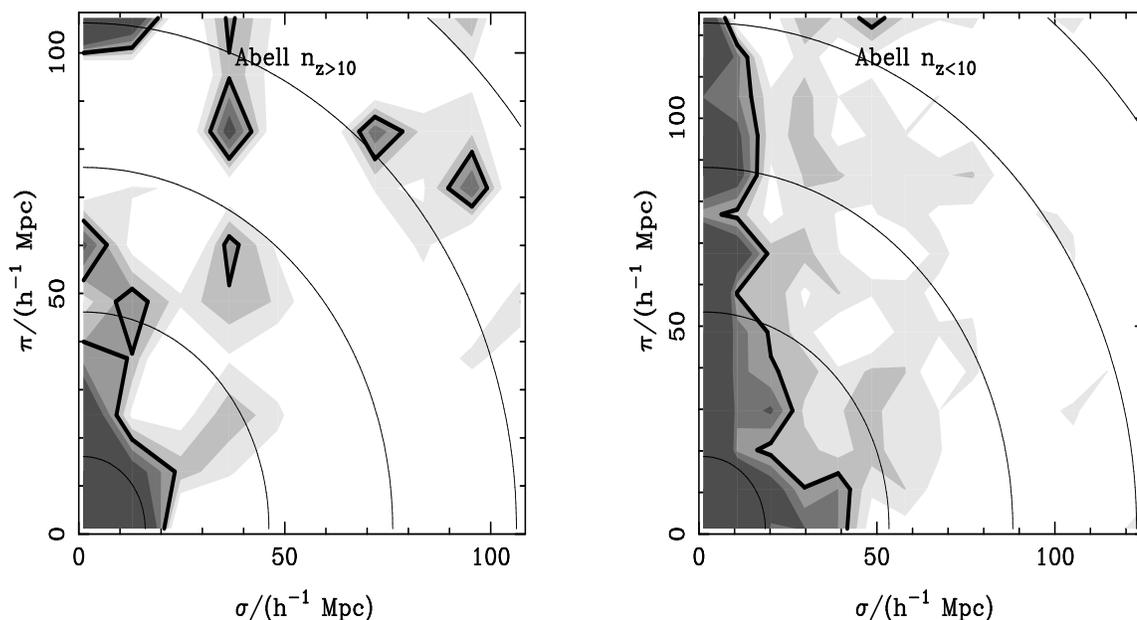}}
\caption{
Correlation function of samples of Abell 
with different number of redshifts used in their distance determinations.  The
left panel shows correlation function contours obtained
from Abell clusters whose distances have been obtained using more than
ten galaxy redshifts.  The results for Abell clusters with less
than ten redshift measurements are shown on the right hand side panel.
Shadings and line conventions are as in figure 1.
}
\label{fig:dist.nz}
\end{figure*}

Figure \ref{fig:dist.nz} shows the results from
clusters whose distances have been obtained using more than
ten galaxy redshifts, $n_z>10$ (left panel), 
and results for clusters with $n_z<10$
(right hand side panel).
The net effect of having a large $n_z$ is to lower the 
amplitude of the line of sight elongation.  It can be seen that, 
even though this effect is not small, it is insufficient to 
reproduce the flattening observed in less massive groups 
of galaxies as seen in figure \ref{fig:dist.str_apm}.
The results shown in the two  panels in figure \ref{fig:dist.nz} 
can be compared to figure 6, paper I,
for different $n_z$ values.
These samples 
were constructed in order to match as closely as possible
the procedure followed to construct cluster samples obtained from angular
information.
By comparison, the trend of increasing anisotropies 
for lower number of redshift measurements used to determine cluster distances is
present in both, the observational and simulated samples.  

An interesting point to be noted is
the clear enhancement of the correlation length in
the $\sigma$ direction for the small $\pi$ bin (which we will
call $\sigma_0$ from now on) in the observational and mock
samples with low $n_z$,
with respect to those with more redshift measurements.
By inspection to figure \ref{fig:wdnz.s0.obs} we see that
this difference is also present in the redshift-space correlation
length $s_0$.  This is in agreement with
more general results obtained from mock cluster catalogues identified from
angular simulated data. This can be seen in figure 10 of Paper I,
where a decrement in $s_0$ for higher values of $n_z$ can be easily seen.
This indicates that it is likely that the difference in
correlation amplitudes indicated by the value of $s_0$
shows the diminished effects of projection
biases in the high $n_z$ Abell sample, rather than
a difference in average cluster mass.

We will come back to this point in the following sections,
where we will examine subsamples of optically and x-ray
selected clusters of galaxies.

\subsubsection{Results from different from Bautz-Morgan type samples}
\label{ssec:opticalbm}

\begin{figure*}
{\epsfxsize=16.truecm \epsfysize=8.truecm 
\epsfbox[33 313 567 561]{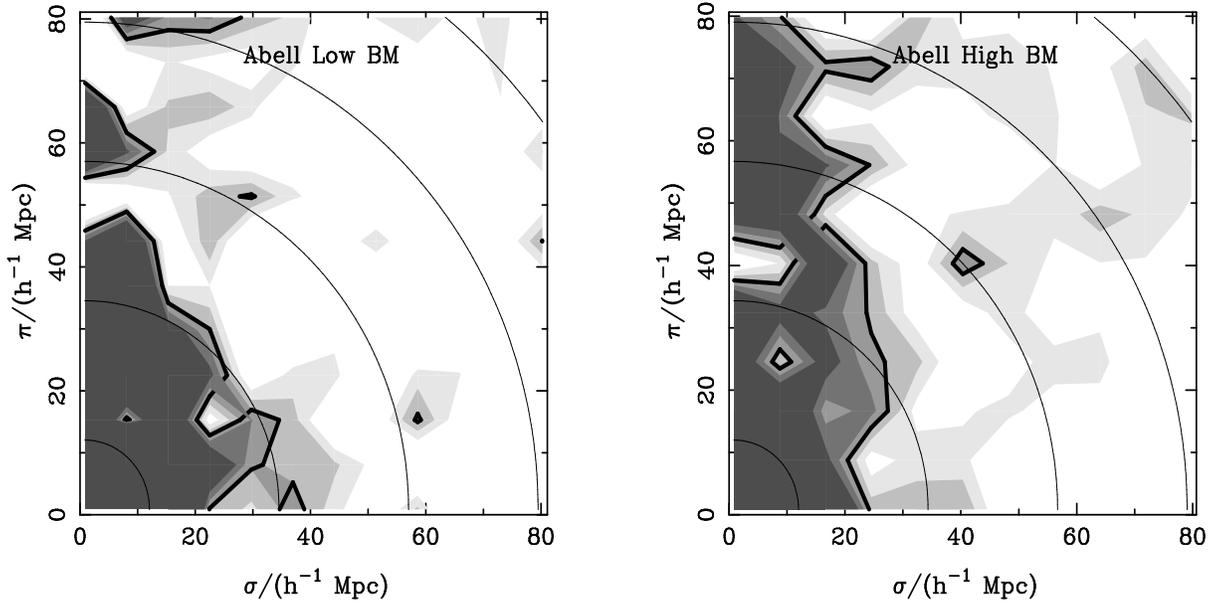}}
\caption{
Redshift-space correlation function for Abell Clusters with different
Bautz-Morgan type.  The left panel shows
the correlation contours for clusters with low Bautz-Morgan type 
(I, I-II, and II).  Results from high Bautz-Morgan types (II-III and
III) are shown in the right hand side panel.
Shadings and line conventions are as in figure 1.
}
\label{fig:dist.bm}
\end{figure*}

We have also 
analysed samples of Abell clusters with different Bautz-Morgan (BM) types.
The BM type is indicative of the regularity of the distribution of galaxies in a
cluster; the larger the BM type, the more irregular the shape of the cluster is.

The results from studying samples of different BM type can be seen in figure 
\ref{fig:dist.bm}, where the left panel shows
the correlation contours for clusters with Bautz-Morgan type 
I, I-II, and II, corresponding to more regular clusters,
and the right panel shows 
results from BM types II-III and III, that is the most irregular ones.
The particular choice of
BM type to define the samples corresponds 
to a rough classification of regular
and irregular clusters.  Even though the differences between both panels
are not conclusive, there is a larger 
elongation in the correlation function of clusters with larger BM types
(the inferred relative velocities for these
cluster samples can be seen in figure \ref{fig:wdnz.s0.obs}).

The larger elongation seen in the
correlation function of high BM type clusters in figure
\ref{fig:dist.bm} indicates that they
are slightly more prone to the inclusion of spurious
background or foreground galaxies than low BM types, 
or even to be results of projection effects
in the angular distribution of galaxies.  These problems seem also
to be present in the low BM type clusters, but to a 
lesser extent.

It can also be appreciated that the estimated correlation length $\sigma_0$
has similar values for the low and high BM type clusters 
(figure \ref{fig:dist.bm}).  
This fact would suggest a similar real-space correlation length for both samples.
However, we expect that the most massive clusters would 
show a more regular galaxy distribution, and therefore, 
be of low BM types, whereas high BM type clusters would correspond to 
less massive clusters, with lower galaxy richness counts, 
more irregular shapes, and lower correlation lengths.  
Since high BM clusters are likely to be subject to larger 
projection effects due to their irregular appearance
they could have a biased large estimate of $\sigma_0$ as shown in 
paper I, clarifying the results of $\sigma_0$ seen in this figure.

We next measure the redshift-space correlation length, $s_0$,
for both sub-samples.  In this case, we
obtain a higher correlation length for the
low BM type clusters (see figure \ref{fig:wdnz.s0.obs}), 
in support of the presence of a correlation
between mass and BM type proposed in the last paragraph. 

Furthermore, as we discussed in the previous sub-section,
it can be argued that 
both $\sigma_0$ and $s_0$
depend on the degree of projection effects present in a sample.
The analyses carried out in Paper I indicate that
the effects of increasing $n_z$ on $s_0$, for samples
of clusters identified from angular data,
is that of making $s_0$ ever smaller and closer to its
actual value.  This result implies that the fact that we obtain
a larger value of $s_0$ for clusters with low BM type 
actually indicates that this sample contains more massive
clusters than that with high BM type clusters.

From these results, we notice that we can reduce the
degree of contamination in a sample of Abell clusters by
considering subsamples of regularly shaped systems.
However, it should be noticed
that projection effects cannot be completely 
removed in the Abell catalogue by considering 
subsamples with low BM type nor large $n_z$.

\subsection{X-ray selected clusters of galaxies}
\label{ssec:xray}

\begin{figure*}
{\epsfxsize=15.truecm \epsfysize=8.truecm 
\epsfbox[33 322 567 561]{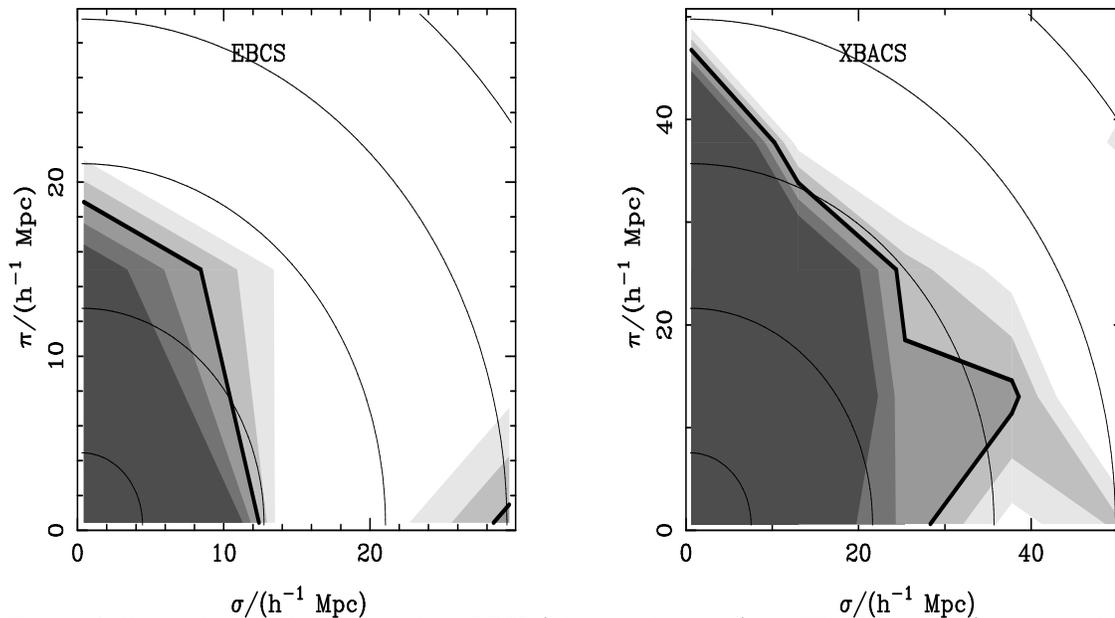}}
\caption{
Correlation function contours 
from EBCS (left hand side panel) and XBACS clusters 
(right hand side panel).
Shadings and line conventions are as in figure 1.
}
\label{fig:dist.xray}
\end{figure*}

In order to provide an insight to the characteristics of cluster samples
with X-ray selection criteria,
we study two samples of X-ray selected
clusters of galaxies, the X-ray Bright Abell Cluster Sample 
(XBACS, Ebeling et al. 1996) 
and the Extended ROSAT Brightest Cluster Sample (EBCS, Ebeling et al. 2000).
The fact that these clusters have been identified from the X-ray
emission of hot gas trapped in the huge potential wells of massive
clusters of galaxies, makes these samples more reliable
than cluster samples selected from 2D data.  This is
a consequence of the fact that the emission from the hot gas is proportional to
the square of the density, which makes it a lot more difficult
to have a spurious cluster of galaxies resulting from the alignment
of small hot gas clouds along the line of sight.

Clusters of galaxies in the XBACS sample, 
which comprises Abell clusters whose emission in X-rays has been confirmed,
would be less affected by projection effects present in the Abell
sample, since only real clusters would show hot gas emission.
On the other hand, in the EBCS sample, sources were identified from
an X-ray survey and then galaxy redshift distributions were analysed 
to find the optical counterparts.  
These clusters should be largely free from projection effects,
and should not include identification biases associated to the Abell cluster
finding algorithm.
However, the reliability of X-ray cluster distances 
obtained from correlations with redshift surveys are subject to uncertainties
that could affect the results. 
The correlation functions measured for the EBCS and XBACS 
samples are shown in the left and right hand side 
panels of figure \ref{fig:dist.xray} respectively.  
As can be seen,
the EBCS sample shows only a slight elongation along the line of sight
which can be easily reconciled with the presence of distance
measurement errors similar to those necessary to explain the
anisotropies observed in the APM correlation function (Padilla \& Baugh, 2002). 
On the other hand, the line of sight distortion of the
XBACS correlation function shown in the
right panel of this figure 
provides a strong indication of the presence of projection effects.
Given their larger correlation amplitude, XBACS should be 
subject to a strong infall signal, which is not observed.
The results from Paper I indicate that
the projection effects seen in the correlation function of
mock clusters identified from angular data
(sample 3) are still present in a set of clusters
with angular positions coincident with
clusters identified in 3D (sample 4).  
The studies performed in Paper I show 
that the elongations in $\xi(\sigma,\pi)$ found in 
this mock cluster catalogue
are quite severe, and are mainly produced by the erroneous use of foreground
and background galaxies in assigning the distances to clusters.

\subsubsection{Luminosity effects in x-ray selected samples}
\label{ssec:xrayluminosity}

\begin{figure*}
{\epsfxsize=15.truecm \epsfysize=8.truecm 
\epsfbox[30 300 570 560]{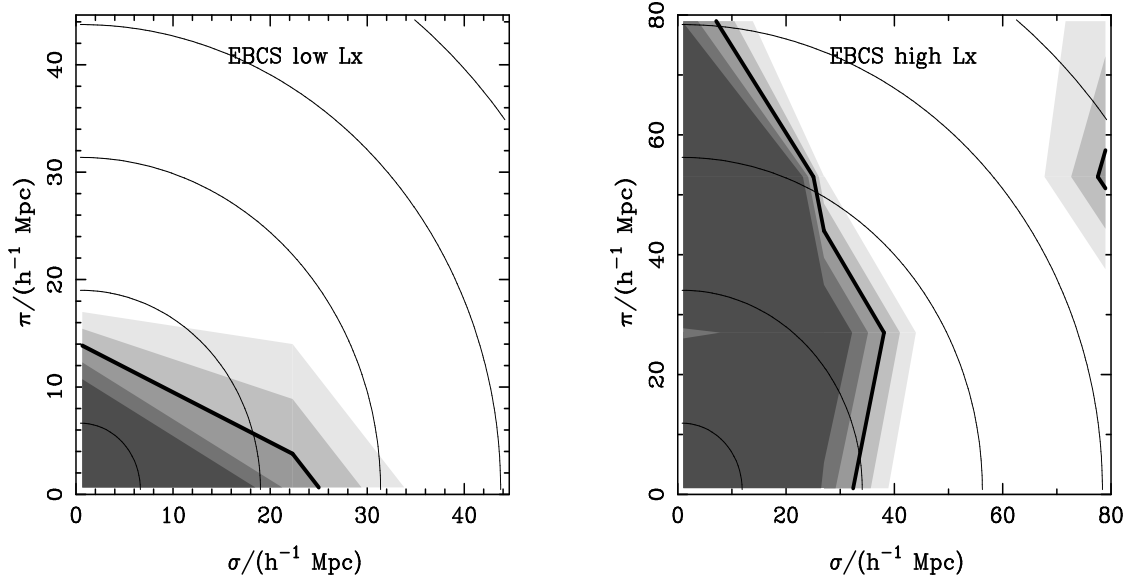}}
\caption{
Correlation function for subsamples of EBCS clusters with different
luminosities.  The left panel shows
the correlation contours for clusters with $L_x<3 \times 10^{37}W$.  
Results from using $L_x>3  \times 10^{37}W$ are shown
in the right hand side panel. 
Shadings and line conventions are as in figure \ref{fig:dist.str_apm}.
}
\label{fig:dist.ebcs.lx}
\end{figure*}

Even though the elongation of the correlation function contours
along the line of sight present in the total EBCS sample 
can be accounted for by redshift measurement
errors, we explore EBCS subsamples defined by limiting the range of
intracluster gas luminosity in x-rays.  We define a limit 
$L_{limit} = 3 \times 10^{37} W$
which divides the total sample of EBCS clusters 
into two subsamples with similar number of clusters.  
The resulting iso-correlation
levels can be seen in figure \ref{fig:dist.ebcs.lx}, where the left panel shows 
the results for $L_x < L_{limit}$, and the right hand side panel, those 
for $L_x > L_{limit}$.  This is a somewhat unexpected result,
since high luminosity clusters show a very elongated pattern, whereas the
least luminous clusters show the expected flattened pattern indicative of
less significant projection effects.

\begin{figure}
{\epsfxsize=8.truecm \epsfysize=8.truecm 
\epsfbox[30 150 560 680]{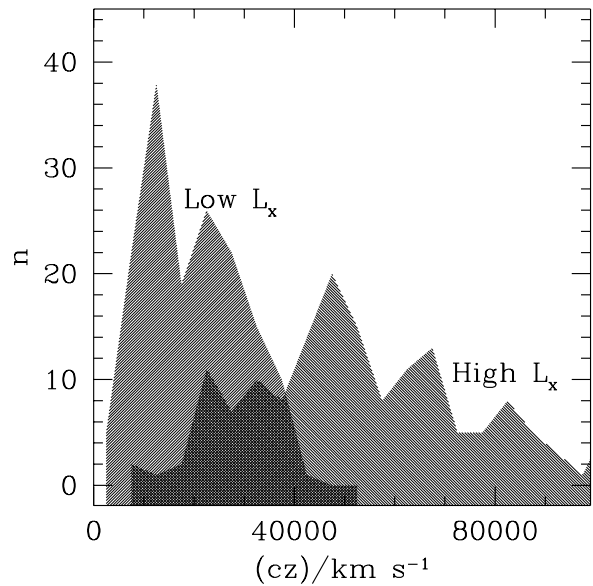}}
\caption{
Redshift distribution for the two EBCS subsamples defined
by a luminosity cut $L_x < L_{limit}$ (upward hatched region), and
$L_{limit} < L_x$ (downward hatched region).
}
\label{fig:nzebcs}
\end{figure}

A possible source for this effect could rely on 
contamination by foreground structures which are
likely to affect more strongly 
clusters at greater distances. These contamination effects could
bias distance estimates and the X-ray luminosity.
In figure \ref{fig:nzebcs} we show the distribution of cluster distances for the
high and the low x-ray luminosity subsamples which support this scenario. 

The resulting values of relative velocities, redshift-space
correlation lengths and bias factors corresponding to
the EBCS subsamples defined by the luminosity cut $L_{limit}$
are presented in figures \ref{fig:wdnz.s0.obs} and \ref{fig:bias.obs}
respectively.  As can be seen, low x-ray luminosity EBCS clusters
present a remarkably low pairwise 
velocity dispersion, comparable to that resulting from the analysis of
UZC groups, identified from redshift data.

\subsubsection{Subsamples of different BM type effects in x-ray clusters}
\label{ssec:xraybm}

\begin{figure*}
{\epsfxsize=15.truecm \epsfysize=8.truecm 
\epsfbox[33 322 567 561]{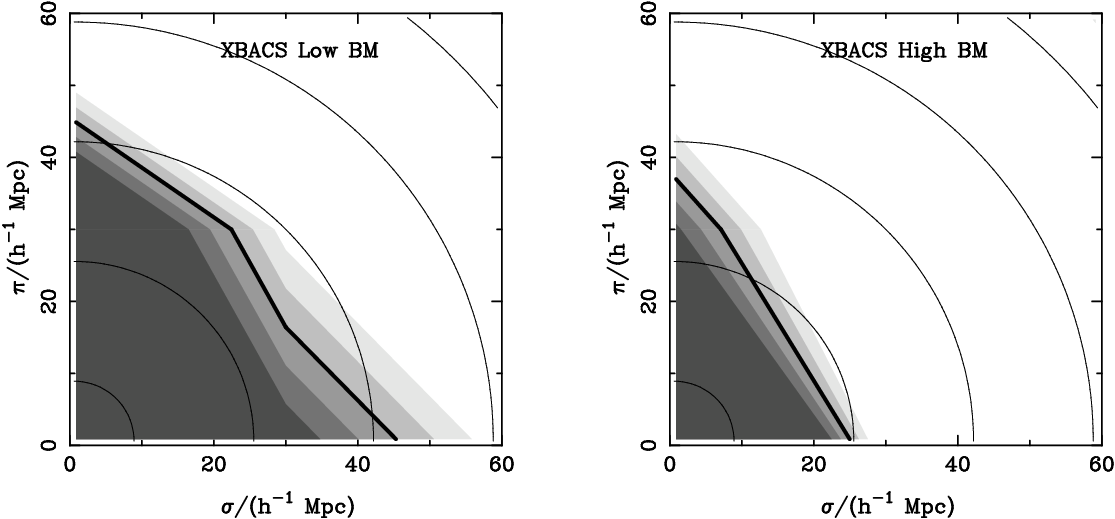}}
\caption{
Correlation function for subsamples of XBACS clusters with different
Bautz-Morgan type.  The left panel shows
the correlation contours for clusters with Bautz-Morgan type I,
I-II and II.  Results from higher Bautz-Morgan types are shown
in the right hand side panel. 
Shadings and line conventions are as in figure \ref{fig:dist.str_apm}.
}
\label{fig:dist.xbacs.bm}
\end{figure*}

In this subsection we analyse the redshift space clustering pattern
arising in samples of x-ray clusters with different BM types.
This is aimed to study the combination of projection effects 
which could manifest in the correlation function.
Since BM types are available for Abell clusters, we restrict the analysis
to the XBACS sample.

\begin{figure}
\epsfxsize=7.truecm \epsfysize=11.truecm 
\epsfbox[40 221 326 795]{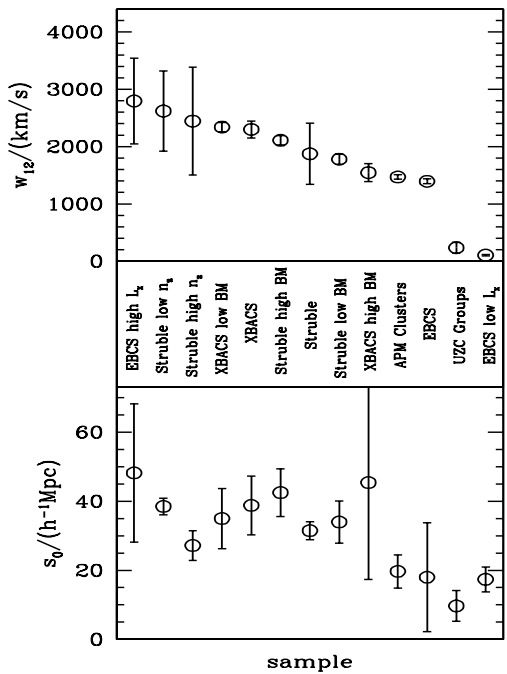}
\caption{
Pairwise velocities (upper panel) and correlation lengths 
(lower panel) obtained for
the different observational samples. 
The samples are distributed along the x-axis, such that their
values of $w_{12}$ decrease to the right of the plot.  This
ordering is maintained in the lower panel.  
}
\label{fig:wdnz.s0.obs}
\end{figure}

\begin{figure}
\epsfxsize=7.truecm \epsfysize=8.truecm 
\epsfbox[55 410 326 695]{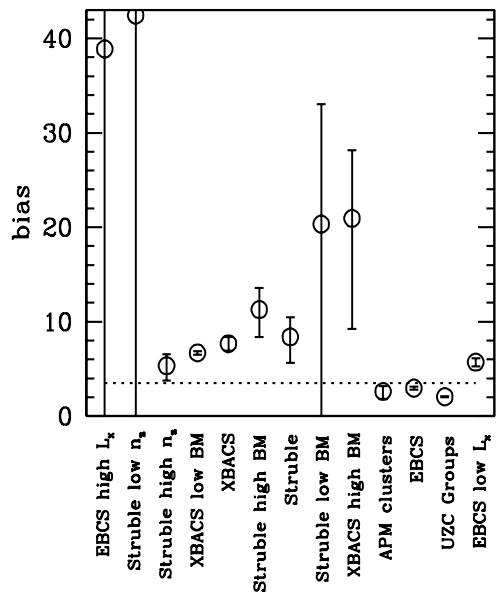}
\caption{
Bias factors for the observational cluster samples studied in this paper.
The circles with errorbars show the results from minimising
equation \ref{eq:chi}, and the solid line shows the predicted 
effective bias from the CDM model and the Sheth, Mo \& Tormen (2001)
mass function.  The range of acceptable values
for the effective bias factors are shown as dotted lines.  
The ordering of the samples is the same as in figure \ref{fig:wdnz.s0.obs}.
}
\label{fig:bias.obs}
\end{figure}

The results can be seen in figure 
\ref{fig:dist.xbacs.bm}, where the left panel shows
the correlation contours for clusters with Bautz-Morgan type 
I, I-II, and II and the right panel shows 
results from BM types II-III and III.
The choice of
BM types for the different panels is the same as that used for
Abell clusters, but the difference in elongations
is more readily apparent.
The figure shows a significantly larger 
elongation for the sample of clusters with larger BM types,
as was the case with Abell clusters of different BM type.
The larger correlation length $\sigma_0$
for the low BM type clusters, which can also be seen as a
larger value of $s_0$ (see figure \ref{fig:wdnz.s0.obs}) can be appreciated
once more.  
This supports the earlier hypothesis that the most massive 
clusters would be more frequently identified
as low BM type clusters, and therefore show a large correlation length.
This figure also supports the idea that the larger elongations
in the high BM type clusters are probably produced by a larger number of
background or foreground galaxies used in assigning
distances to high BM type clusters.

\subsection{Relative velocities and
correlation lengths in observational cluster samples}
\label{ssec:w12obs}

In order to summarise the results obtained from the study of observational
samples of clusters of galaxies identified from angular data, 
we present the values of
relative velocities and redshift-space correlation lengths
obtained from the correlation functions shown in previous figures.

We find the relative velocities by minimising
equation \ref{eq:chi}.  The real space correlation function used in this
equation corresponds to a CDM power spectrum with parameters
$\Omega_{matter}=0.3$, $\Omega_{\Lambda}=0.7$,
a CDM shape parameter $\Gamma=0.2$, and $\sigma_8=0.9$.  

Based on the results of
the study of galaxies and groups of galaxies, and supported by
simulations of 
the hierarchical processes thought to drive the formation of structure,
we expect to find smaller relative velocities for clusters of galaxies
than are found for galaxies and groups.  
The upper panel of figure \ref{fig:wdnz.s0.obs}
shows relative velocity on the y-axis, and observed sample on the x-axis.
The errors in relative velocity result from the different estimates
obtained from different correlation function levels, and show
1-$\sigma$ confidence levels.  
The ordering of the samples along the
x-axis is such that relative velocities decrease to the right.
If the cluster samples were free from projection effects of any 
kind, this ordering
of samples would result in showing groups on the
left and the most massive clusters on the right.  However, as can be seen,
every cluster sample, with the exemption of EBCS clusters
with low x-ray luminosity, is located to the left of this plot, showing their
large contamination due to projection effects.  The lower panel shows
the redshift-space correlation length with errorbars obtained
from performing a power law fit to the measured redshift-space
correlation functions over a range of scales dependent on the
sample analysed $r_{min}<r<r_{max}$, where 
$2.5 < r_{min}/$h$^{-1}$Mpc$ < 10$, and
$28 < r_{max}/$h$^{-1}$Mpc$ < 80$
(the ordering of samples in this
panel is the same as in the previous one).  There are signs
of a weak correlation between the amplitude of the relative velocities and 
the correlation length for the cluster samples.  
According to the results from Paper I, the degree of projection effects
correlates with both, the value of the relative velocities and the
correlation length,  so a similar trend in both parameters was
expected; the deviations from this behaviour reflect the different
minimum mass and X-ray flux 
thresholds in the different observational samples, which
in the case of the mock samples with different $n_z$ was always
the same by construction.

In a hierarchical scenario of structure formation it is expected that
cluster relative velocities should not be larger than galaxy pairwise velocities
($\sim 400 km/s$). The observations
suggest apparent large pairwise velocities 
$w_{12}$, an effect 
which could have a large contribution
from redshift measurement
errors, and even more important contributions from large systematics
due to projection biases present in some of the samples
identified  from angular data. 

  This is the case with Abell clusters
and the XBACS sample, except when 
considering subsamples of clusters with low BM type.
Large relative velocities are inferred for luminous EBCS clusters,
giving hints for the presence of contamination due to projection effects.
However, the results for EBCS 
of low x-ray luminosity, show significantly smaller pairwise velocities
comparable to those found for the UZC groups, and consistent with
the predictions of a hierarchical clustering scenario.

\subsection{Bias factors}
\label{ssec:biasobs}

The comparison between the measured and predicted redshift-space
correlation function can be used to obtain estimates of
the bias parameter of the cluster sample under consideration,
using equation \ref{eq:chi}. 
In figure \ref{fig:bias.obs} 
we show the different bias parameters obtained 
for the different samples analysed.  The errors on the bias
parameter come from the different results obtained from
the comparison of $5$ levels of correlation function.

These values can be compared to what is expected from a particular
CDM model, using the expression for an effective bias as presented in
Padilla \& Baugh (2002), which is a weighted average of the
bias of a sample of clusters of mass $M$, derived
by Sheth, Mo \& Tormen (2001), and requires knowledge of the space
density of the cluster sample.  The results from using
this equation are  valid if the cluster sample
is complete above a minimum mass threshold.

We have derived approximate estimates of the number density of 
clusters by
assuming that the cluster samples are complete above
a minimum cluster mass limit.  This rough approximation only
applies to optically selected clusters since x-ray selected 
samples are flux-limited with a minimum mass which strongly depends on 
redshift.
Then we calculate the space density of optical 
cluster samples by simply counting the number of objects
up to a distance where the 
redshift distribution departs from the $r^2$ behaviour due to the influence of 
a selection function. 
The effective
CDM bias for the measured cluster number densities are then 
derived straightforward following Padilla \& Baugh (2002).
Given the uncertainties in the assumptions
mentioned before, and the large scatter between cluster richness and 
virial mass, we just provide a suitable upper limit
for the bias parameter, namely the largest value obtained, $b=3.5$, 
corresponding to the Abell sample with low BM type
(this is equivalent to a mean intercluster separation $d_c\simeq 60$h$^{-1}$Mpc
or a number density $n\simeq 5 \times 10^{-6}$h$^3$Mpc$^{-3}$).
This upper limit is shown as a dotted line in figure \ref{fig:bias.obs}.

For some of the samples studied, the resulting bias parameters
obtained from the $\chi^2$ minimisation 
are well above our estimated upper limit (figure \ref{fig:bias.obs}).
However, UZC groups, APM clusters, EBCS total sample and 
low x-ray luminosity subsample, and Abell clusters with high $n_z$,
show lower values of the bias parameter,
compatible with our expectations.

\section{An insight on the results from groups and clusters identified
from redshift data}

The UZC group sample was constructed by identifying 
galaxy systems from the 3D distribution of galaxies in redshift space from 
the Updated Zwicky catalogue.
By inspection of figure \ref{fig:wdnz.s0.obs},
the great leap between the results of UZC groups 
relative velocities, and those obtained from samples
of clusters identified from angular data, either in the
optical or x-ray wavelength range, is clear.  This is also apparent in figure 
\ref{fig:dist.str_apm}, where the anisotropies present in the correlation
function of the UZC groups show markedly flattened iso-correlation contours, 
a pattern we almost completely failed to  
reproduce by measuring the correlation function of other cluster samples 
throughout this paper with the exception of the low
x-ray luminosity EBCS subsample.

In this section, we explore in more detail the
relative velocities, redshift-space correlation lengths and bias parameters
obtained from the correlation function of UZC groups as a function
of group richness counts. We remark the difference between
the number of member galaxies in a group and the values of $n_z$ from
the previous analysis, for in this case the number of galaxies is
associated to the identification process and not only to the number of redshift
measurements in the field of a cluster. 

\begin{figure}
\epsfxsize=8.truecm \epsfysize=9.truecm 
\epsfbox[30 160 556 665]{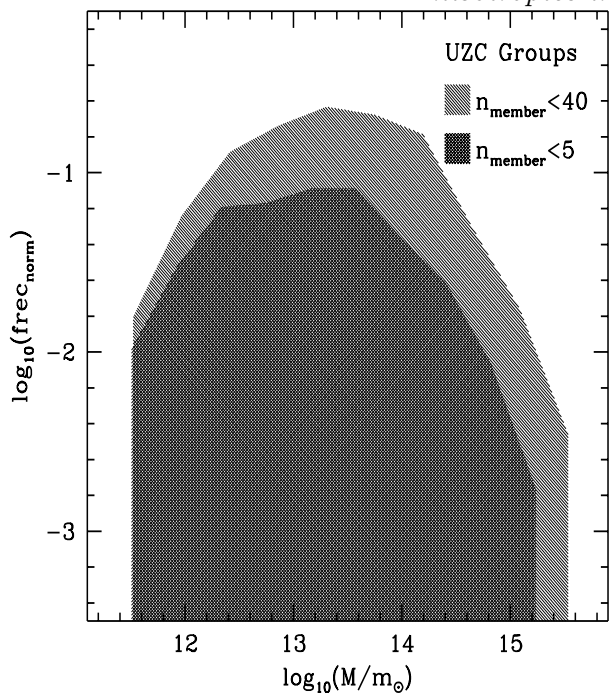}
\caption{
Virial masses distribution function for the UZC groups, for different
subsamples, defined by an upper limit in $n_{member}$.  The light
shaded area corresponds to $n_{member}<40$, and the dark shaded area to
$n_{member}<5$.
}
\label{fig:mass.uzc}
\end{figure}

We note that the number
of member galaxies is correlated to some extent with the mass of the groups.
This can be seen in figure \ref{fig:mass.uzc},
where we show the distribution of group virial masses for two different
upper limits in number of member galaxies.  The normalisation constant
by which both distribution functions were multiplied, is 
set up so as to make the integral of the
mass distribution function of groups with $n_{member}<5$ equal to $1$.
It can be seen that the
sample with a low upper limit for $n_{member}$ 
(dark shaded area) shows a marked relative
lack of high mass groups when compared to the sample with a high
upper limit in member galaxies (light shaded area).  One of the
effects this leads to is the intrinsic change in the clustering
amplitude of the different $n_{member}$ 
subsamples.  It should be noted that samples with different
$n_z$ in mock sample 2, are not affected by this bias.  This is readily
apparent when recalling a result from Paper I, where it was shown that
this sample is consistent with a constant redshift-space correlation 
length for any value of $n_z$.

\begin{figure}
\epsfxsize=7.truecm \epsfysize=12.truecm 
\epsfbox[55 180 350 695]{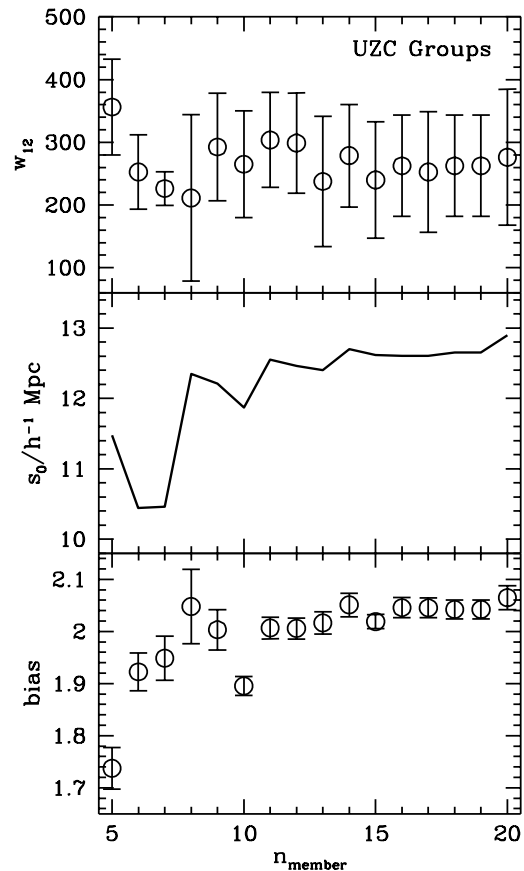}
\caption{
UZC groups
relative velocities (upper panel), redshift-space correlation lengths
(middle panel), and bias factors (lower panel) as a function of minimum
number of member galaxies.  
}
\label{fig:uzc}
\end{figure}

Following the procedures given in sections \ref{ssec:w12obs} and
\ref{ssec:biasobs}, we calculate the relative velocities,
redshift-space correlation functions, and bias corresponding
to subsamples of the UZC groups with different number of 
member galaxies. The results can be seen in figure \ref{fig:uzc}.
The upper panel shows values for the relative velocities which 
are consistent with the published value of $w_{12}=250 \pm 110$km/s
(Padilla et al. 2001) for any value of $n_{member}$.  There are also
signs of a somewhat larger value of $w_{12}$ for small $n_{member}$,
but within the size of the errorbars.  The middle panel shows the
values of $s_0$ as a function of $n_{member}$.  As can be seen, there
is a tendency for higher values of $s_0$ as the number of member galaxies
increases.  This increase is also present in the measurements of bias
factors, shown in the lower panel, and is consistent with the results
from figure \ref{fig:mass.uzc}, which show an increasingly
larger average mass for samples with larger $n_{member}$.  The higher
the average mass of a sample, the higher its correlation amplitude
and bias factor will be.  
The increase in $s_0$ and bias factor are only
noticeable for $n_{member} \leq 10$.  Most of
the samples with a larger number of member galaxies show
stable and constant results for the three parameters in study,
$w_{12}$, $s_0$ and bias factor, in good agreement with
the results obtained from mock sample 2, which also showed this
behaviour.
The errorbars in figure
\ref{fig:uzc} shown for $w_{12}$ and
bias factor are obtained from the different results
obtained by using five correlation function amplitudes ($\xi=0.6, 0.8, 1.0,
1.2,$ and $1.4$).

The main conclusion of this section is the remarkable 
stability of the results for the UZC groups. 
This provides a clear indication that redshift space identification 
of clusters is a powerful procedure to obtain samples
with no significant contamination. 
The weak dependence of the results on a wide range of number of 
member galaxies reflects
the reliability of the identification procedure.

\section{Conclusions}

We have explored
the correlation function of clusters in redshift space
for different observational samples of galaxy clusters. 
We have studied the consequences of
different observational issues on the distortion of the clustering pattern, 
such as different sources of cluster identification, number of redshift
measurements, cluster morphology and X-ray luminosity.

Padilla et al. (2001) showed that the results from
groups of galaxies identified from 3-dimensional information
available in the UZC, are consistent with the
hierarchical clustering paradigm which predicts that  
we should be able to find
objects that are still falling towards more massive structures.  This
can be observed in the anisotropy of the correlation function contours
of the UZC groups, which shows the flattening produced by this infall
motion.  This is not easily detectable for galaxies 
since they are in a strongly non-linear
regime and the pairwise peculiar velocities dominate the dynamics. 

The automatic cluster finding algorithms used in the identification
of APM clusters, make them a more reliable sample of clusters, which
have been shown to be largely free of projection effects.  
Although contamination can not be totally excluded, the
inclusion of an error
in the APM cluster distance measurement is sufficient to account for
for the elongations found in the APM clusters redshift-space correlation
function (Padilla \& Baugh 2002).  

On the other hand,
Abell clusters show clear signs of projection effects (this work,
Sutherland 1988; Sutherland \& Efstathiou 1991; see also Lucey 1983),
which can be reduced by restricting the sample of clusters
in richness class (Miller et al. 1999), for example.  In this work,
we found that the elongations in the correlation function isopleths
can be reduced when restricting the sample to have a low BM type,
which is a measure of the regularity of the
galaxy distribution in a cluster.  We
have also shown that this could be related to an increase in
the richness of the clusters considered, since our 
correlation function measurements show an increase in
the correlation length. This makes sense, given that it is
reasonable to suppose that richer clusters of galaxies have a more 
regular shape.
In the end, the anisotropies left after the restriction of the Abell
sample cannot be easily reconciled with errors in distance measurement
from using a small number of galaxies.  The results indicate the presence
of galaxies not physically bounded to the cluster, and 
also a fraction of spurious clusters resulting from structures
along the line of sight.

This situation is not significantly improved in the case of
X-ray clusters.  Even though the XBACS sample 
is composed of Abell clusters 
with confirmed emission in X-rays, they also show significant elongations
along the line of sight when compared to the results from
the Abell sample.  We found that restricting
the sample of XBACS clusters to have a low BM type
produces a decrease in the elongation of the
iso-correlation contours along the line of sight.

On the other hand, EBCS clusters, which are identified primarily as
X-ray sources, show an interesting behaviour that deserves further attention.
Low X-ray luminosity clusters, $L_x< 3 \times 10^{37}$W
 show little correlation function
anisotropy indicating the lack of significant projection effects. 
Clusters of high X-ray luminosity, 
on the other hand, exhibit strong line of sight distortions,
indicating the presence of large biases in this subsample. 
It can be argued that luminous clusters are more distant and
therefore most strongly affected by contamination effects that could
bias distance and x-ray luminosity estimates. 

The results shown in figures \ref{fig:wdnz.s0.obs} and \ref{fig:bias.obs},
which correspond to observational datasets identified from angular data, 
indicate that the values of relative velocities
obtained from cluster samples are usually well over the
values preferred by hierarchical clustering.  It is also noticeable the
large variations in redshift-space correlation lengths which, as
shown in Paper I, is also influenced by the level of contamination present
in the sample.  Those observational samples less affected by
projection effects, show values of the bias parameter 
compatible with a suitable upper limit 
inferred using a number density threshold $n=5 \times 10^{-6}$h$^3$Mpc$^{-3}$
(equivalent to a mean intercluster separation $d_c\simeq 60$h$^{-1}$Mpc),
and a $\Lambda$CDM model in combination with the Sheth, Mo \& Tormen
(1999) mass function.  

With the advent of  surveys of the next generation,
the construction of new group and cluster catalogues from deep
spectroscopic data based on accurate photometry, 
will provide more reliable and well controlled cluster samples.
In the light of our results for presently available samples, we 
find that detailed analysis of the redshift-space correlation 
function anisotropies can give a deep insight on sample characteristics
and invaluable information on their implications.

\section*{Acknowledgments}
This work  was supported in part by CONICET, Argentina,
and the PPARC rolling grant at the University of Durham.  
DGL acknowledges support from the John Simon Guggenheim Memorial Foundation.
We thank the Referee for invaluable comments and advice
which greatly improved the previous version of the paper.
We have benefited from helpful discussions with Carlton Baugh.
We acknowledge the Durham Extragalactic Astronomy Group and the 
Virgo Consortium for making the Hubble Volume 
simulation mock catalogues available.


\begin{thebibliography}{}
\bibitem[]{10}
Abadi, M.G., Lambas, D.G., Muriel, H., 1998, ApJ, 507, 526.%
\bibitem[]{30}
Bahcall, N.A., Cen, R.Y., Gramann, M., 1994, ApJ, 430, L13.%
\bibitem[]{40}
Bahcall, N.A., Soneira, R.M., 1983, ApJ, 270, 20.%
\bibitem[]{50}
Bahcall, N.A., Soneira, R.M., Burgett, W.S., 1986, ApJ, 311, 15.%
\bibitem[]{55}
Bautz, L.P. \& Morgan, W.W., 1970, ApJ, 162, 149.%
\bibitem[]{80}
Borgani, S., Plionis, M., Kolokotronis, V., 1999, MNRAS, 305, 866.%
\bibitem[]{100}
Cole, S., Hatton, S.J., Weinberg, D.H., Frenk, C.S., 1998, MNRAS, 300, 945.%
\bibitem[]{130}
Collins, C.A., Guzzo, L., Boehringer, H., Schuecker, P, Chincarini, G., Cruddace, R., De Grandi, S, MacGillivray, H.T., Neumann, D.M., Schindler, S., Shaver, P., Voges, W., 2000, MNRAS, 319, 939.%
\bibitem[]{150}
Croft, R.A.C., Efstathiou, G., 1994, MNRAS, 268, L23.%
\bibitem[]{160}
Croft, R.A.C., Dalton, G.B., Efstathiou, G., Sutherland, W.J., Maddox, S.J., 1997, MNRAS, 291, 305.%
\bibitem[]{170}
Dalton, G.B., Efstathiou, G., Maddox, S.J., Sutherland, W.J., 1992, ApJ, 390, L1.%
\bibitem[]{180}
Dalton, G.B., Efstathiou, G., Maddox, S.J., Sutherland, W.J., 1994, MNRAS, 269, 151.%
\bibitem[]{190}
Dalton, G.B., Maddox, S.J., Sutherland, W.J., Efstathiou, G., 1997, MNRAS, 289, 263.%
\bibitem[]{210}
Davis, M., \& Peebles, P.J.E., 1983, ApJ, 267, 465.%
\bibitem[]{230}
Ebeling, H.; Voges, W.; Bohringer, H.; Edge, A. C.; Huchra, J. P.; Briel, U. G., 1996, MNRAS, 281, 799.%
\bibitem[]{240}
Ebeling, H.; Edge, A. C.; Allen, S. W.; Crawford, C. S.; Fabian, A. C.; Huchra, J. P., 2000, MNRAS, 318, 333.%
\bibitem[]{255}
Eisenstein, D., \& Hu, W., 1998, ApJ, 496, 605.%
\bibitem[]{270}
Evrard, A. E.; MacFarland, T. J.; Couchman, H. M. P.; Colberg, J. M.; Yoshida, N.; White, S. D. M.; Jenkins, A.; Frenk, C. S.; Pearce, F. R.; Peacock, J. A.; Thomas, P. A.  2002, ApJ, 573, 7.%
\bibitem[]{275}
Huchra, J.P., \& Geller, M.J., 1982, ApJ, 256, 423.%
\bibitem[]{290}
Lucey, J. R., 1983, MNRAS, 204, 33.%
\bibitem[]{300}
Lumsden, S.L., Nichol, R.C., Collins, C.A., Guzzo, L., 1992, MNRAS, 258, 1.%
\bibitem[]{315}
Miller, C.J., Krughoff, K.S., Batuski, D.J. \& Hill, J.M., 2002, AJ, 124, 1918.%
\bibitem[]{320}
Miller, C.J., Batuski, D.J., Slingend, K.A., Hill, J.M., 1999, ApJ, 523, 492.%
\bibitem[]{325}
Netterfield, C. B. et al., 2002, ApJ, 571, 604.%
\bibitem[]{328}
Padilla, N.D., \& Lambas, D.G., 2003, MNRAS, submitted (Paper I).%
\bibitem[]{330}
Padilla, N.D., \& Baugh, C.M., 2002, MNRAS, 329, 431.%
\bibitem[]{340}
Padilla, N.D., Merchan, M.E., Valotto, C.A., Lambas, D.G., Maia, M.A.G., 2001, ApJ, 554, 873.%
\bibitem[]{350}
Peacock, J. A. \& West, M.J., 1992, MNRAS, 259, 494.%
\bibitem[]{360}
Postman, M., Huchra, J.P., Geller, M.J., 1992, ApJ, 384, 404.%
\bibitem[]{}
Sheth, R.K., Mo, H.J., Tormen, G., 2001, MNRAS, 323, 1.%
\bibitem[]{375}
Struble, M.F. \& Rood, H.J., 1999, ApJS, 125, 35.%
\bibitem[]{380}
Sutherland, W., 1988, MNRAS, 234, 159.%
\bibitem[]{390}
Sutherland, W.J., Efstathiou, G., 1991, MNRAS, 248, 159.%
\bibitem[]{400}
van Haarlem, M.P., Frenk, C.S., White, S.D.M., MNRAS, 1997, 287, 817.%
\end{thebibliography}
\end{document}